# The Impact of the Virtualization of Scholarly Conferences on the Gender Structure of Conference Contributors

Agnieszka Olechnicka* (ORCID: 0000-0001-5525-636X)
Adam Ploszaj* (ORCID: 0000-0002-6638-3951)
Ewa Zegler-Poleska* (ORCID: 0000-0002-7574-5985)

* University of Warsaw, Science Studies Lab & Centre for European Regional and Local Studies EUROREG, Warsaw, Poland

Corresponding author: Adam Płoszaj, a.ploszaj@uw.edu.pl

Version: 28-10-2024

**Abstract**

The underrepresentation of women in academic conferences is an underexplored aspect of gender disparity in science. This study aims to expand knowledge on this issue by investigating whether the virtualization of academic conferences in the wake of the COVID-19 pandemic changed the gender structure of conference participants. We explored this question utilizing authorship data from the Web of Science Conference Proceedings Citation Index for 180 conferences in 30 conference series held between 2017 and 2023, with a total number of 88,384 papers and 404,295 authors. At least one edition of each analyzed conference series was launched in a virtual or hybrid form. This sample enables a comparison of differences in the gender participation of conference authors while controlling for heterogeneity among conference series. Using linear and logistic regression models, we identified a positive difference in women's involvement in virtual and hybrid conferences compared to onsite events. However, this effect was due less to the increased participation of women in virtual and hybrid conferences than to the decreased participation of women in the onsite editions of the analyzed conference series.

## 1. Introduction

Gender disparity in academia remains evident. Women generally publish fewer papers, achieve lower citation rates, and face more challenges in disseminating their research (Song et al., 2024; Sugimoto et al., 2015; Vásárhelyi et al., 2021). Despite the gradual increase in the representation of women in research staff and grant awards, the gender gap in research output has persisted over the past two decades (Elsevier, 2024). The underrepresentation of women extends to academic conferences, where women remain less visible, particularly as speakers and in high-status roles (Blumen & Bar-Gal, 2006; Braun et al., 2023; Falk & Hagsten, 2022). This further undermines their position because scholarly conferences are critical for scientific discussion, communication, and networking – essential components for career development (Hansen & Budtz Pedersen, 2018; Jacobs & McFarlane, 2005; Leon & McQuillin, 2020; Teplitskiy et al., 2024).

The COVID-19 pandemic significantly disrupted scholarly activities. It led to a surge in journal publications, albeit with unequal impacts across academic groups (García-Costa et al., 2024; Ioannidis et al., 2022; Kwon et al., 2023; Madsen et al., 2022). The literature confirms the gender gap in number of publications across disciplines (Jemielniak et al., 2023), the intensity of new project initiation (Gao et al., 2021), and available time for scientific work (Esquivel et al., 2023; Myers et al., 2020).

At the same time, the shift to virtual platforms transformed scholarly communication and networking practices (Waltman et al., 2021). The impact of virtual conferences on gender inequality remains a subject of debate, with mixed findings (Olechnicka et al., 2024). For example, while certain studies indicate positive changes in gender composition among conference participants due to the online transition (Biermann, 2024; Skiles et al., 2022; Walton et al., 2022), while others report that gender disparities persist or even worsen (Falk & Hagsten, 2022; Jarvis et al., 2023; Standaert & Thunus, 2022; Zhang et al., 2023). These inconsistent findings highlight the need for comprehensive studies that analyze gender participation over a more extended period and across multiple conference formats, using reliable and objective data of high credibility.

This study seeks to address this gap by exploring how the virtualization of academic conferences due to the COVID-19 pandemic has influenced the gender structure of participants. The study addresses the research question: How has the virtualization of academic conferences due to the COVID-19 pandemic altered the gender structure of conference participants? The hypothesis tested in this paper is that virtual and hybrid modes of academic conferences are associated with higher participation of women scholars than onsite conferences.

Conducting research in this area is crucial to assess the scale of changes in women's participation in academic conferences due to shifts in their format and highlight virtualization as a potential driver for increasing women's presence in academia. This study contributes to the broader discourse on the positive impact of gender diversity on scientific innovation, research productivity, and the visibility of academic work (Hofstra et al., 2020; Mulders et al., 2024).

The paper is structured as follows. It begins with a review of the literature on gender disparity in academia particularly in scholarly conferences, and its evolution during the COVID-19 pandemic. Detailed descriptions of the data sampling, methods, and results follow. The paper finishes with a discussion and conclusions drawn from the study.

## 2. Literature Review

However, the share of female researchers globally has approached parity (47% in 2022), and 37% of research grants are now awarded to women, the gender gap in research output and outreach as well as career progression persists (Astegiano et al., 2019; Elsevier, 2024; Halevi, 2019). Women generally publish fewer papers and have lower citation rates than men, but these differences vary by field and country (Boekhout et al., 2021; Huang et al., 2020). Men continue to dominate fields like mathematics, physics, and software engineering, while women are more productive and highly cited in the female-dominated field of nursing but not in psychology (Sá et al., 2023). Additionally, there remains significant underrepresentation of women in STEM disciplines and patent creation (Elsevier, 2024). Furthermore, altmetric studies indicate that women are less successful in disseminating their research online (Song et al., 2024; Vásárhelyi et al., 2021). Factors contributing to this disparity include differences in career length, dropout rates, institutional support, and stereotypes relating to gender (Duch et al., 2012; Huang et al., 2020; Jadidi et al., 2018; Nielsen et al., 2017). Country-level development indicators also play a role. Lower human development and higher gender inequality correlate with reduced female participation in academia (Elsevier, 2024; Larivière et al., 2013).

**Gender Disparity in Scholarly Conferences**
The issue of gender disparity extends beyond research output and recognition. Studies of participation in conferences and other academic meetings consistently reveal a persistent underrepresentation of women, particularly in prestigious roles. This disparity is evident across various disciplines, highlighting the widespread nature of the problem. As one example, despite increased overall participation, women remain underrepresented in high-status positions at the annual meetings of the Israeli Geographical Society (Blumen & Bar-Gal, 2006). This trend is also evident in the fields of academic tourism, hospitality, leisure, and events, where gender inequality persists among keynote speakers and members of honorary committees (Walters, 2018). In evolutionary biology, and gambling studies women are less frequently represented as invited speakers at conferences, However, the proportion of invited women tends to be higher when there are more women among the organizers (Débarre et al., 2018; Monson et al., 2023). Similarly, gender and geography significantly influence author representation at the Association for Information Science and Technology Annual Meeting (Buchanan & McKay, 2022). In critical care, despite some progress, gender disparity remains, with male speakers still outnumbering female speakers at conferences (Dymore-Brown et al., 2024). Women are also underrepresented in computer science as colloquium speakers at top universities, but having female colloquium chairs increases the likelihood of female speakers (Nittrouer et al., 2018). On average, women accounted for less than one-third of the speakers at medical conferences held between 2017 and 2018 in Australasia, Canada, Europe, the UK, and the US, but the proportion varied significantly, ranging from 5.8% to 74.5% (Arora et al., 2020). In economic conferences, paper submissions authored solely by women are 3.3% less likely to be accepted, largely due to gender stereotypes in the review process (Hospido & Sanz, 2021).

The findings discussed above can be disturbing because academic conferences are essential for scientific work and scholarly communication, serving as vital spaces in which scientific knowledge is scrutinized, debated, and refined (Hansen & Budtz Pedersen, 2018; Jacobs & McFarlane, 2005). Gender disparities in participation and recognition at these events not only impede equitable career advancement for women but also limit the diversity of perspectives that are critical for driving innovation in science and scientific progress (Hofstra et al., 2020; Mulders et al., 2024; Nielsen et al., 2017). Addressing this imbalance is essential to ensuring conferences foster inclusivity and maximize the potential of the scientific community.

Finally, it should be noted that gender disparities intersect with geographic location and socio-economic conditions, creating additional barriers to participation for scholars from underrepresented regions (Mickey & Smith-Doerr, 2022). Women from countries with lower gender equality indices or fewer economic resources may face compounded challenges, even in virtual formats, due to digital exclusion and lack of institutional support. Efforts to enhance inclusivity at conferences must address not only gender but also these broader intersectional factors (Kozlowski et al., 2022). However, as a recent review (Robinson-García et al., 2024) highlights, advances in scientometric tools enable nuanced analyses of diversity, yet many aspects, including socio-economic background, are difficult to measure.

**Gender Disparity in Scientific Productivity Amidst COVID-19**
Academic work has been disrupted by the COVID-19 pandemic. The increasing demand for research due to the COVID-19 pandemic and anti-contagion public measures impacted academic productivity positively. Interestingly, although the abnormal peak of submissions was dominated by health and medical researchers during the early stages of the pandemic, submissions to social science and economics journals later increased (García-Costa et al., 2024). However, the so-called "covidisation of research" did not affect all academic groups equally, and the citation impact for COVID-19 publications far exceeded that of works in other fields, significantly influencing the citation profiles of scientists and shaping the scientific elite (Ioannidis et al., 2022).

In addition, the positive impact was not equal. During the first wave of the pandemic, junior women scientists submitted proportionally fewer manuscripts than men (García-Costa et al., 2024; Madsen et al., 2022). This was even more prominent for junior women working in less prestigious academic organizations located in less gender-equal countries (Kwon et al., 2023). Even though some studies did not observe a decline in the number of manuscripts submitted by women, they indicated that the pre-pandemic rise in manuscript submissions had subsided and that authors in all journals, countries, and fields were overwhelmingly men before, during, and after the pandemic (Son & Bell, 2022).

Significant differences in publication patterns between genders were reported in different disciplines, although there were no significant differences between men and women in overall publication patterns between 2019 and 2021 (Jemielniak et al., 2023). The pandemic negatively impacted many female scholars' work habits and routines due to higher competing demands from family obligations, such as home-schooling and parental care (Esquivel et al., 2023; European Commission: Directorate-General for Research and Innovation, 2023). Female scientists, especially those with young children, experienced a decline in the rate of new project initiation (Gao et al., 2021) and a substantial drop in time devoted to research (Myers et al., 2020).

**The Virtualization of Scientific Conferences and the Gender Gap**
Scholarly communities responded to pandemic-related restrictions by moving research communications to the virtual space (Waltman et al., 2021). Virtual conferences have been lauded for their inclusiveness, allowing women, early-career scientists, people with disabilities, and researchers from less affluent countries to participate by reducing costs and eliminating travel barriers. Still, the transition to remote or hybrid formats introduced several challenges that limited potential inclusiveness, such as digital exclusion, reduced networking opportunities, and lower conference engagement of attendees with caregiving responsibilities, particularly women (Olechnicka et al., 2024).

Several empirical studies reported reduced gender disparity resulting from transitioning scholarly conferences to an online format. For instance, the transition to a virtual mode in 2020 for three scientific conferences—the International Conference on Learning Representations (ICLR), the American Astronomical Society (AAS), and the North American Membrane Society (NAMS)—increased the number of attendees compared to traditional in-person meetings, with higher rises in female participation (60–260%) than in male participation (Skiles et al., 2022). Similarly, the shift to remote presentations for economics seminars held by 270 institutions worldwide between 2018 and 2021 led to a significant increase in the number of women leading seminars (Biermann, 2024). The shift to online formats during the pandemic further increased women's participation, even in conferences where they were already highly represented, such as the Agriculture, Nutrition & Health Academy Week in 2021, where women made up 65% of attendees (Walton et al., 2022), and the International Communication Association, where female participation rose from 52% in 2005 to 59% in 2022 (Braun et al., 2023).

Nonetheless, other studies indicate that virtual meetings have exacerbated gender inequalities in academic settings. For instance, an online survey distributed to 542 academic researchers from five Belgian universities one month into the COVID-19 lockdown revealed that the overall number of meetings increased from 5.50 to 6.08 per week during the pandemic. Although men experienced a significant increase in meetings, the increase for women was not substantial, widening the gender gap. Additionally, 31% of women, particularly women in lower hierarchical ranks, reported greater difficulty speaking up in virtual meetings compared to 20% of men (Standaert & Thunus, 2022). Several studies also highlight the persistence of gender inequalities in virtual academic environments, but they can serve more as background points than primary references for the proposed research. For instance, Jarvis and colleagues (2023) found that men continued to dominate question-and-answer sessions in virtual conferences; Zhang, Torchet, and Julienne (2023) observed that although gender parity was achieved in the audience, women asked half as many questions as men; Falk and Hagsten (2022) found that the facilitation of participation through the online format did not significantly increase the proportion of women among keynote speakers.

The findings from recent studies indicate that while virtual conferences have lowered entry barriers and increased women's access to scientific events, they have not resolved the gender disparities in active participation and visibility. The rise in female participation in virtual conferences in quantitative terms has not been accompanied by a corresponding improvement in the qualitative dimension, such as delivering keynote lectures or engaging actively in discussions. This indicates that, despite the potential of online conferences to facilitate greater inclusivity, structural gender inequalities remain deeply entrenched and require further investigation and targeted interventions.

This study addresses the existing research gaps by contributing to the understanding of the long-term effects of virtual and hybrid conference modes on gender equality in academia. While previous studies have often focused on short-term impacts or individual events during the peak of the pandemic, this study provides a comprehensive analysis that examines gender participation over a longer period and across multiple conference formats, using reliable and objective data of high credibility. This approach offers valuable insights for advancing the discussion on gender equality in academic participation.

## 3. Data and Methods

Whereas prior studies primarily relied on conference registration data (Skiles et al., 2022; Zhang et al., 2023), conference programs (Falk & Hagsten, 2022; Jarvis et al., 2023), survey results (Standaert & Thunus, 2022), in-depth interviews, focus groups, and video recordings of virtual events (Walton et al., 2022), our approach utilizes a different source: conference proceedings indexed in the Web of Science database (WoS, Clarivate). In May 2024, we searched for conference proceedings indexed in the Conference Proceedings Citation Index (CPCI), including both the Conference Proceedings Citation Index – Science (CPCI-S) and the Conference Proceedings Citation Index – Social Science & Humanities (CPCI-SSH), which we accessed through the University of Warsaw Library. Our searches were deliberate and targeted, rather than random, as the latter would have been unfeasible. We aimed to identify international conferences from different disciplines, occurring regularly, and resulting in proceedings with data enabling a gender analysis of the authors' names. The resulting set of conferences does not meet the criteria for a truly random sample. However, we believe that it avoids biases other than those resulting from the database's characteristics.

In the first step, we limited the search in WoS CPCI to the years 2017–2023 to collect proceedings data covering seven years, with the pandemic year 2020 serving as a midpoint. Next, we inspected the search results using the "Conference Titles" filter and further refined the results using other available WoS filters. We used "Conference Titles" and not "Publication Titles" because the proceedings of different conferences may be published in the same publication series (such as Lecture Notes in Computer Science). A notable issue we encountered was inconsistency in conference titles. For example, the titles of the conference editions varied slightly, so the proceedings were indexed under several different titles rather than one consistent title.

In the next step, we focused on the conferences whose titles seemed relevant to our study, specifically those indicating that the conferences were held regularly (titles containing words such as "annual meeting") and had an international scope ("international conference"). We then checked whether proceedings were available for 2017–2023. We aimed to cover the entire timespan, but it proved difficult. Finally, we inspected the details of the proceedings to verify whether the full first names were provided. If the verification was positive, we included the conference for analysis and downloaded the proceedings' full records from WoS as Excel files. Nevertheless, we also encountered inconsistencies in WoS data because some editions of the same conference did not feature full first and last names.

We collected data for all document types and did not exclude any type of document indexed in CPCI. However, we decided early on to exclude large medical conference proceedings for two main reasons. First, we aimed to cover various disciplines in our data. Second, on a more practical level, medical conference proceedings tended to provide only the initials of the authors' first names rather than the full names, making it impossible to conduct a gender analysis. We acknowledge that such conferences might warrant a separate study due to their specific characteristics and the existing literature addressing gender issues in the field of medicine. Overall, this process resulted in the selection of 30 conference series encompassing 180 individual events, in which 88,384 papers were presented, with a total number of 404,295 authors. The list of the full names of the conferences selected for the analysis is presented in Table A1 in the Appendix. Additionally, Table A2 in the Appendix shows disciplines (defined as Web of Science research areas) covered by each selected conference.

We defined four conference modes: onsite, virtual, hybrid, and switched. The onsite mode is the traditional meeting in a physical location where participants gather in person, while the virtual mode is an online event that allows participants to attend from anywhere in the world. The hybrid mode combines the two previous modes. Finally, the switched mode refers to a meeting that was initially planned as onsite but was subsequently "switched" to a virtual one. In our study, the switched mode occurred only in 2020 and is analyzed together with the onsite mode because participants registered assuming that it would be onsite.

Web of Science provides information on the locations of conferences, including online events, designated as "ELECTR NETWORK." However, it does not include information on hybrid events. To address this gap, we conducted internet searches for each edition of each conference, consulting the official conference websites and other available sources, such as calls for papers published on external websites, to verify the location and assign a mode to the event. We encountered some difficulties in classifying editions into a specific mode. For instance, the website of the 2023 edition of the Conference on AI, Ethics, and Society (AIES), stated that "the virtual option is not for presenting authors," so we classified it as an "onsite" event. In another example, the International Conference on Artificial Neural Networks (ICANN) website indicated that its 2020 conference had been canceled, but the proceedings were published and appeared in WoS CPCI; thus, we classified this edition as "onsite."

Table 1 presents the abbreviated names of the conferences and their modes. Blank cells indicate that no proceedings data was available in the Web of Science database. With proceedings data for only 15 conferences, 2023 features the most gaps, 2021 was dominated by the virtual mode, and the hybrid mode prevailed in 2022. Overall, after the virtual boom in 2021, there was a gradual return to the onsite mode.

**Table 1. Modes of the analyzed conference series**

| Conference series abbreviation | 2017 | 2018 | 2019 | 2020 | 2021 | 2022 | 2023 |
|---|---|---|---|---|---|---|---|
| ACHEMS |   | o | o |   | v | v | o |
| ACL | o | o | o | v | v | h | h |
| AEA | o | o | o | o | v | v | o |
| AIED | o | o | o | s | h | h |   |
| AIES |   | o | o | o | v | h | o |
| BIOPHYS | o | o | o | o | v | o |   |
| CHI | o | o | o | s | v | h | h |
| COMPNET | o | o | o | s |   | h | o |
| CSEDU | o |   | o | s | v | v |   |
| DGO | o | o | o |   |   | v | o |
| ECGBL | o | o | o |   | v |   |   |
| ECKM | o | o | o | s | h |   |   |
| EDUCON | o | o | o | s | v | h |   |
| EUCAP | o |   | o | s | v | h | o |
| GLOBECOM | o | o | o | v | h | h |   |
| HEAD | o | o | o | s | v | h |   |
| ICANN | o | o | o | o | v | h | h |
| ICIP | o | o | o | s | v | h | h |
| ICML |   | o | o | v | v | h |   |
| IJCNN | o | o | o | s | v | o | o |
| INTER | o | o | o | s | h | h |   |
| IUS | o | o | o | s | v | h |   |
| JCDL | o | o | o |   | v | h |   |
| KDD | o | o | o | s | v | o | o |
| MME | o | o | o | s | h | h |   |
| NENE | o | o | o | h | h |   |   |
| ROMAN | o | o | o | s | v | h | h |
| SMART | o | o | o | v | v |   |   |
| WACV | o | o | o | o | v | o | o |
| WEBCONF | o | o | o | s | v | h | o |

Legend

| | |
|---|---|
| o | onsite |
| s | switched |
| v | virtual |
| h | hybrid |

For the purposes of this research, we treat gender as a binary variable, including only men and women, but we acknowledge that there are other gender identities, such as non-binary or trans (Lindqvist et al., 2021). The collected proceedings data was processed using a custom R script to extract authors' first and last names and affiliations. The gender of the conference contributions' authors was identified based on their first and last (family) names as well as the country of their academic affiliation. To do so, we used NamSor gender detection tool. The effectiveness of this tool has been verified in independent empirical tests, the results of which indicate a very high accuracy of the tool compared to other available solutions (Sebo, 2021). A gender was assigned to 99.7% of the authors of the analyzed conference contributions.

The number of women authoring conference papers between 2017 and 2023 was relatively stable, with the average percentage of women authors ranging from 29.4% to 34.1%. In the analyzed set, an upward trend in the share of women among the authors of the analyzed conference contributions is visible (Table 2). At the same time, there is also a very large variation in the participation of women in individual conferences, which is reflected in the large difference between the minimum and maximum values (full data on the participation of women in the individual editions of the analyzed conference series are presented in Table A5 in the Appendix).

**Table 2. Share of women among the authors of the analyzed conference contributions (%)**

|                    | 2017 | 2018 | 2019 | 2020 | 2021 | 2022 | 2023 |
|--------------------|------|------|------|------|------|------|------|
| Average            | 29.4 | 30.3 | 31.2 | 30.7 | 34.1 | 33.4 | 33.0 |
| Standard deviation | 9.0  | 8.4  | 9.8  | 8.5  | 9.3  | 8.4  | 7.4  |
| Min                | 13.0 | 14.6 | 12.6 | 14.4 | 16.9 | 18.0 | 16.6 |
| Max                | 54.0 | 54.4 | 60.7 | 54.4 | 56.4 | 56.7 | 45.8 |

The analysis carried out in this study had two variants associated with different ways of calculating the explained variable: OLS and Logit. In the first analysis, the explained variable was the percentage of women authors in the proceedings of the analyzed conferences. Because this is a continuous variable, the appropriate analytical method is simple linear regression. The second approach estimated the probability that a presentation's author was a woman. Since this variable is binomial, the appropriate analytical method is logistic regression. The main difference between the two approaches is the level of analysis and the related number of observations. In the first approach, the level of analysis is the conference; in the second approach, it is the conference participant. Performing an analysis at the level of conference participants allowed us to control the covariates describing individual participants. In this analysis, we considered the country of affiliation of the conference participant and the participant's country GDP per capita (which was impossible in the analysis at the conference level).

**Table 3. Variables employed in the analyses**

| Type of variable      | Variable                                                                                      | OLS | Logit |
|-----------------------|-----------------------------------------------------------------------------------------------|-----|-------|
| Explained variables   | Share of women among the authors of conference contributions                                  | ✓   |       |
|                       | Gender of the authors of conference contributions (1 = women, 0 = men)                        |     | ✓     |
| Explanatory variable  | Conference mode (1 = virtual or hybrid; 0 = onsite or switched from onsite to virtual/hybrid) | ✓   | ✓     |
| Control variables     | Conference year                                                                               | ✓   | ✓     |
|                       | Conference series (dummy)                                                                     | ✓   | ✓     |
|                       | Number of papers presented (log)                                                              | ✓   | ✓     |
|                       | Conference host country (dummy)                                                               | ✓   | ✓     |
|                       | Participant country (dummy)                                                                   |     | ✓     |
|                       | Participant's country GDP per capita (log)                                                    |     | ✓     |

In both analyses, the primary variable of interest was the type of conference, a binary variable taking the value of 1 for conferences held in virtual or hybrid mode and 0 for conferences held in onsite mode or planned as onsite but changed to virtual or hybrid mode (switched). The switched type appeared only in the pandemic year of 2020. The control variables used in both analyses were conference series, year of conference organization, host country of the conference, and the total number of papers presented at a given conference. Moreover, the Logit analysis used a country of the participant affiliation and the participant's country GDP per capita to account for between country differences. In both the OLS and Logit variants, several specifications were estimated with a smaller or larger range of control variables (for details, see the results section).

Importantly, individual editions in the conference series could not be treated as independent observations, primarily because regular conferences usually attract the same participants and have their own rules and customs—all of which may affect the explained variable we are interested in. To mitigate this presumed observation dependence, the model used robust errors clustered by conference series. This approach allowed us to estimate the correct standard errors and significance levels despite violating the requirement of independence of observation.

## 4. Results

Between 2017 and 2019, the participation of women in analyzed onsite conferences slightly increased, from 29.4% to 31.2%. The year 2020 stands out because women's participation in onsite and switched conferences was at 32.4%, maintaining the previous trend, while their participation in virtual and hybrid conferences was low, around 23.4%.. However, 2020 was unique in that only five conferences in our dataset were categorized as virtual or hybrid because most events had been planned as onsite and were switched at the last minute. The shift to entirely virtual and hybrid conference formats after 2020 had a positive impact on women's participation, which reached 34.1% in 2021 and 2022, and 35.2% in 2023. In 2022 and 2023, a noticeable difference in women's participation between onsite and virtual or hybrid conferences was observed: 34.1% vs. 29.3% in 2022 and 35.2% vs 31.8% in 2023 (see Figure 1). This suggests that the gender disparity in conference participation was primarily driven by a decrease in women's participation in onsite conferences rather than an increase in virtual conference participation.

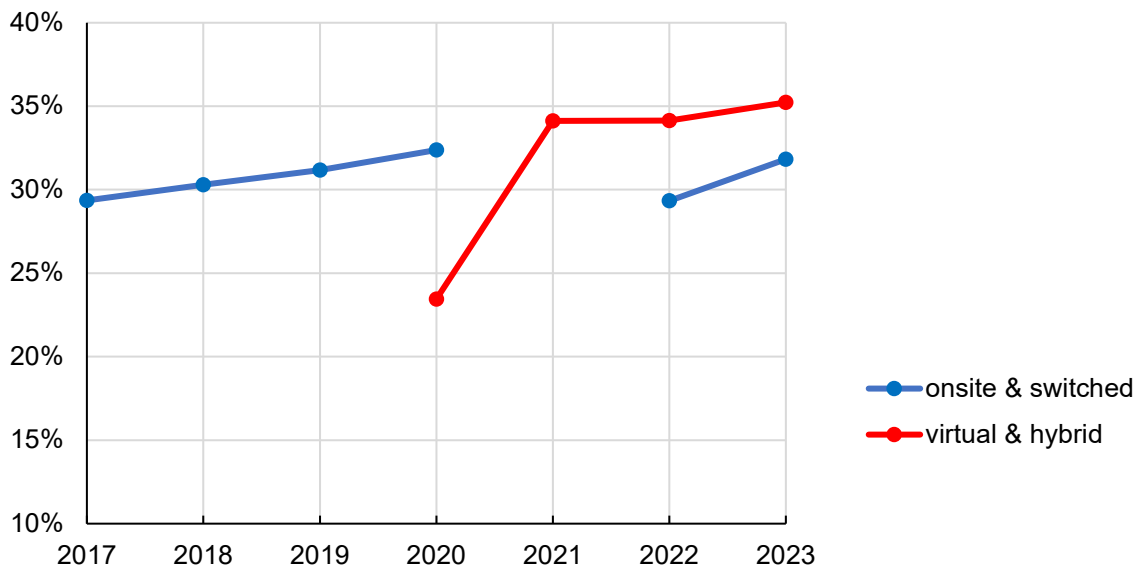

**Figure 1. Percentage of women among the authors of conference contributions by conference mode**

The results of the statistical modeling are presented in Tables 4 and 5. Table 4 displays the results of the linear model (OLS), in which the explained variable was the percentage of women among the authors of conference contributions. The analysis was performed for two specifications that differ only in including the host country dummy (i.e., information about the country where the conference was held). Both specifications had a very high value of the R-squared statistic, which resulted from the use of a conference series dummy. Both specifications provided consistent results indicating a relationship between conference mode and the percentage of women participating. The effect had quite good statistical significance (p-value = 0.016 in Specification 1 and p-value = 0.02 in Specification 2), especially considering the relatively small sample (180 observations). At the same time, this observed effect, although statistically significant, was relatively modest in its scale. The parameter estimates of 0.013 in Specification 1 and 0.021 in Specification 2 mean that the expected difference in the percentage of women among the authors of conference contributions between onsite conferences and virtual or hybrid conferences was circa 1.3 percentage points (Specification 1) or 2.1 percentage points (Specification 2).

**Table 4. Effect on the percentage of women among the authors of conference contributions, OLS results**

|                           | (1)       | (2)       |
|---------------------------|-----------|-----------|
| Virtual or hybrid         | 0.013**   | 0.021***  |
|                           | (0.005)   | (0.007)   |
| Conference year           | 0.006***  | 0.006***  |
|                           | (0.002)   | (0.002)   |
| Number of papers (log)    | 0.028     | 0.029     |
|                           | (0.018)   | (0.021)   |
| Conference series (dummy) | ✓         | ✓         |
| Host country (dummy)      |           | ✓         |
| Observations              | 180       | 180       |
| R-squared                 | 0.924     | 0.939     |

Errors in parentheses (robust, clustered by conference series).
Significance levels: *** $p < 0.01$, ** $p < 0.05$, * $p < 0.1$.

In the variant using logistic regression, the variable was explained by the gender of the conference contribution author, operationalized as 1 for women and 0 for men. The results show the relationship between the examined independent variables and the probability that the gender of the conference participant was a woman. Modeling was performed for four specifications differing in whether the host country dummy and participant country dummy were included. This analysis used a large number of observations: 356,089 for Specifications 1 and 3, and 356,056 for Specifications 2 and 4. The smaller number of observations in Specifications 2 and 4 was the result of the inclusion of the participant country dummy variable. In the case of several countries, this resulted in perfect separation, so these observations had to be excluded for the maximum likelihood estimation algorithm to work correctly. Excluding these observations should not significantly impact the analysis because they constitute only 0.01% of the initial sample.

All presented specifications produced similar results. The virtual or hybrid conference mode (including switched conferences in 2020) was associated with a statistically significant increase in the probability that the author of a conference contribution was a woman compared to the onsite mode. The effect was estimated at 1,043–1,095. The statistical significance of the result was high: 0.001 for Specification 1, 0.013 for Specification 2, below 0.000 for Specification 3, and 0.002 for Specification 4. It is worth emphasizing that specifications including the host country dummy variable provided higher statistical significance of the examined independent variable and indicated a more substantial effect (parameter estimate of 1.071–1.095 in Specifications 3 and 4 compared to an estimate of 1.043–1.054 in Specifications 1 and 2).

**Table 5. Effect on the likelihood that the author of a conference contribution is a woman, Logit results, odds ratio**

|                             | (1)        | (2)        | (3)        | (4)        |
|-----------------------------|------------|------------|------------|------------|
| Virtual or hybrid           | 1.054***   | 1.043**    | 1.095***   | 1.071***   |
|                             | (0.017)    | (0.018)    | (0.026)    | (0.023)    |
| Conference year             | 1.032***   | 1.025***   | 1.034***   | 1.028***   |
|                             | (0.006)    | (0.004)    | (0.005)    | (0.004)    |
| Number of papers (log)      | 1.049*     | 1.023      | 1.045*     | 1.021      |
|                             | (0.030)    | (0.022)    | (0.026)    | (0.021)    |
| GDP per capita (log)        | 0.857***   | 0.819**    | 0.858***   | 0.814**    |
|                             | (0.025)    | (0.075)    | (0.025)    | (0.078)    |
| Conference series (dummy)   | ✓          | ✓          | ✓          | ✓          |
| Participant country (dummy) |            | ✓          |            | ✓          |
| Host country (dummy)        |            |            | ✓          | ✓          |
| Observations                | 356,089    | 356,056    | 356,089    | 356,056    |

Errors in parentheses (robust, clustered by conference series).
Significance levels: *** p < 0.01, ** p < 0.05, * p < 0.1.

The logistic regression results translated into predicted probabilities, specifically the predicted difference in the percentage of women participants between onsite and virtual conferences, largely align with the OLS result. Predicted differences based on logistic models vary between 0.9 and 1.9 percentage points (1.1 for specification, 0.9 for specification 2, 1.9 for specification 3, and 1.4 for specification 4). As we remember, the OLS values were 1.3 and 2.1 percentage points, depending on the specification. The major difference is that logistic regression results have noticeably higher statistical significance statistics, mostly because logistic models are based on a much larger number of observations than the OLS models.

Moreover, it is worth noting that in all presented specifications, there was an observable relationship between the year of the conference and the participation of women. Over the years—even in the short seven-year period we studied—there has been a gradual, although slight, increase in the share of women among the authors of conference contributions. These results were statistically significant in all specifications, both in OLS and logistic models.

## 5. Discussion and Conclusions

In this article, we analyzed the relationship between conference mode and the participation of women as authors in scientific conferences in 2017–2022. Our results demonstrate that the virtual and hybrid conference mode is associated with a greater involvement of women scientists than the onsite mode. Although this effect is statistically significant in our sample, it must be emphasized that its scale is not very large, between circa 1 and 2 percentage points. Furthermore, the distribution of raw data suggests that this should be attributed not to the increased participation of women in virtual and hybrid conferences but to their decreasing participation in onsite editions of the analyzed conference series.

These findings imply that transitioning to virtual and hybrid environments could be an effective strategy for enhancing gender inclusivity in academic conferences, which supports findings from other recent studies (Braun et al., 2023). However, it is important to consider that although virtual and hybrid formats may increase participation, they could also diminish the networking opportunities facilitated by face-to-face interactions during onsite events. This may make virtual events and virtual participation in hybrid events less beneficial for career development, knowledge exchange, and the establishment of collaborative relationships (Campos et al., 2018). It should also be emphasized that participation in virtual conferences requires appropriate infrastructural resources, which may result in the emergence of other dimensions of inequality in scientific communication, perhaps less in the context of gender and more in the context of access to resources and financing. Another factor that may have unintended negative consequences is time zone differences and the resulting inconvenience for peripheral centers having to adjust to the conference time set by privileged centers.

Furthermore, the collected data indicate that virtual scientific conferences may have seen a one-off boom and bust during the pandemic. In 2021, all editions of the analyzed conference series were held in virtual or hybrid mode (with a predominance of virtual); in 2022, the hybrid mode dominated, but several conferences were held in the onsite mode. However, in 2023, there was a clear return to the onsite mode. This may suggest that we are witnessing a return to pre-pandemic patterns of gender participation in conferences. Thus, repeating our study in the future may be impossible due to a lack of virtual conferences. From the point of view of the possibility of conducting further research, this is an unfavorable situation. Nonetheless, the negative aspects of virtual academic conferences (Olechnicka et al., 2024) and virtual fatigue (Nesher Shoshan & Wehrt, 2022) should be acknowledged.

Our analysis is based on a relatively small sample, and a larger sample size would be beneficial. However, our research approach assumed that we analyzed only cyclical conferences whose individual editions were held in different modes. This made it possible to analytically control the features of individual conference series and, as a result, separate the effect that may be attributed to the change in conference mode from other features. It would be worth repeating this analysis in the future, including at least another year of the conference series we analyzed, even if they all return to the onsite mode.

Our study addresses the need for tools to monitor gender gaps at academic conferences (Corona-Sobrino et al., 2020). This work should be continued in several dimensions, one of the most promising of which would be to consider the conference participants' roles. Our analysis encompassed all authors of conference proceedings without differentiation based on their specific roles. Future research could benefit from examining the position of authors in the conference contribution byline, such as first or last authors, to better understand the roles of men and women as lead authors and any other potential changes over time. This approach could provide deeper insights into gender disparity in conference contribution authorship and help identify further avenues for research.

## Statements and Declarations

**Acknowledgments**
We would like to thank Eliasz Czułada, also known as Ilya Alipau, for his contribution to developing the algorithm for processing Web of Science data and determining the gender of authors.

**Author contributions**
Conceptualization: AP, AO, EZP.
Methodology: AP.
Software: AP, EZP.
Formal analysis: AP.
Data curation: AP, EZP.
Writing – original draft: AO, AP, EZP.

Writing – review & editing: AP, EZP, AO.
Funding acquisition: AO.

**Data availability**
This work uses proprietary Web of Science data provided by the University of Warsaw Library.

**Funding**
The University of Warsaw supported this publication under Priority Research Area V of the "Excellence Initiative—Research University" program.

Mickey, E. L., & Smith-Doerr, L. (2022). Gender and innovation through an intersectional lens: Re-imagining academic entrepreneurship in the United States. *Sociology Compass*, *16*(3), e12964. https://doi.org/10.1111/soc4.12964

Monson, E., Ng, K., Sibbick, H., Berbiche, D., & Morvannou, A. (2023). Gender disparity in prestigious speaking roles: A study of 10 years of international conference programming in the field of gambling studies. *PLOS ONE*, *18*(6), e0286803. https://doi.org/10.1371/journal.pone.0286803

Mulders, A. M., Hofstra, B., & Tolsma, J. (2024). A matter of time? Gender and ethnic inequality in the academic publishing careers of Dutch Ph.D.s. *Quantitative Science Studies*, *5*(3), 487–515. https://doi.org/10.1162/qss_a_00306

Myers, K. R., Tham, W. Y., Yin, Y., Cohodes, N., Thursby, J. G., Thursby, M. C., Schiffer, P., Walsh, J. T., Lakhani, K. R., & Wang, D. (2020). Unequal effects of the COVID-19 pandemic on scientists. *Nature Human Behaviour*, *4*(9), 880–883. https://doi.org/10.1038/s41562-020-0921-y

Nesher Shoshan, H., & Wehrt, W. (2022). Understanding "Zoom fatigue": A mixed-method approach. *Applied Psychology*, *71*(3), 827–852. https://doi.org/10.1111/apps.12360

Nielsen, M. W., Alegria, S., Börjeson, L., Etzkowitz, H., Falk-Krzesinski, H. J., Joshi, A., Leahey, E., Smith-Doerr, L., Woolley, A. W., & Schiebinger, L. (2017). Gender diversity leads to better science. *Proceedings of the National Academy of Sciences*, *114*(8), 1740–1742. https://doi.org/10.1073/pnas.1700616114

Nittrouer, C. L., Hebl, M. R., Ashburn-Nardo, L., Trump-Steele, R. C. E., Lane, D. M., & Valian, V. (2018). Gender disparities in colloquium speakers at top universities. *Proceedings of the National Academy of Sciences*, *115*(1), 104–108. https://doi.org/10.1073/pnas.1708414115

Olechnicka, A., Ploszaj, A., & Zegler-Poleska, E. (2024). Virtual academic conferencing: A scoping review of 1984–2021 literature. Novel modalities vs. long standing challenges in scholarly communication. *Iberoamerican Journal of Science Measurement and Communication*, *4*(1), 1–31. https://doi.org/10.47909/ijsmc.93

# Appendix to:

# The Impact of the Virtualization of Scholarly Conferences on the Gender Structure of Conference Contributors

Agnieszka Olechnicka* (ORCID: 0000-0001-5525-636X)
Adam Ploszaj* (ORCID: 0000-0002-6638-3951)
Ewa Zegler-Poleska* (ORCID: 0000-0002-7574-5985)

* University of Warsaw, Science Studies Lab & Centre for European Regional and Local Studies EUROREG, Warsaw, Poland

Table A1. Conferences selected for the analysis

| Short name | Full name |
| --- | --- |
| ACHEMS | Association for Chemoreception Sciences |
| ACL | Annual Meeting of the Association for Computational Linguistics |
| AEA | Annual Meeting of the American Economic Association |
| AIED | International Conference on Artificial Intelligence in Education |
| AIES | AAAI ACM Conference on AI, Ethics, and Society |
| BIOPHYS | Annual Meeting of the Biophysical Society |
| CHI | ACM CHI Conference on Human Factors in Computing Systems |
| COMPNET | International Conference on Complex Networks and their Applications |
| CSEDU | International Conference on Computer Supported Education |
| DGO | Annual International Conference on Digital Government Research |
| ECGBL | European Conference on Games Based Learning |
| ECKM | European Conference on Knowledge Management |
| EDUCON | IEEE Global Engineering Education Conference |
| EUCAP | European Conference on Antennas and Propagation |
| GLOBECOM | IEEE Conference on Global Communications |
| HEAD | International Conference on Higher Education Advances |
| ICANN | International Conference on Artificial Neural Networks |
| ICIP | IEEE International Conference on Image Processing |
| ICML | International Conference on Machine Learning |
| IJCNN | International Joint Conference on Neural Networks |
| INTER | International Speech Communication Association Annual Conference INTERSPEECH |
| IUS | IEEE International Ultrasonics Symposium |
| JCDL | ACM IEEE Joint Conference on Digital Libraries |
| KDD | ACM SIGKDD Conference on Knowledge Discovery and Data Mining |
| MME | International Conference on Mathematical Methods in Economics |
| NENE | International Conference Nuclear Energy for New Europe |
| ROMAN | IEEE International Conference on Robot & Human Interactive Communication |
| SMART | IEEE International Smart Cities Conference |
| WACV | IEEE CVF Winter Conference on Applications of Computer Vision |
| WEBCONF | The Web Conference |

**Table A2.** Conferences and their research areas (as determined in the Web of Science)

| Short name | Research areas |
| --- | --- |
| ACHEMS | Behavioral Sciences; Food Science & Technology; Neurosciences & Neurology; Physiology |
| ACL | Computer Science; Linguistics |
| AEA | Business & Economics |
| AIED | Computer Science; Education & Educational Research |
| AIES | Computer Science; Social Sciences - Other Topics |
| BIOPHYS | Biophysics |
| CHI | Computer Science; Computer Science; Robotics |
| COMPNET | Computer Science; Mathematics; Mathematical Methods In Social Sciences |
| CSEDU | Computer Science; Education & Educational Research |
| DGO | Computer Science; Public Administration; Social Issues |
| ECGBL | Computer Science; Education & Educational Research |
| ECKM | Business & Economics |
| EDUCON | Education & Educational Research; Engineering |
| EUCAP | Engineering; Telecommunications |
| GLOBECOM | Engineering; Telecommunications |
| HEAD | Education & Educational Research |
| ICANN | Computer Science; Computer Science; Radiology, Nuclear Medicine & Medical Imaging |
| ICIP | Computer Science; Engineering; Imaging Science & Photographic Technology; Computer Science; Engineering; Imaging Science & Photographic Technology |
| ICML | Computer Science; Computer Science; Engineering |
| IJCNN | Computer Science |
| INTER | Computer Science; Engineering |
| IUS | Engineering; Engineering; Physics, Acoustics; Engineering; Acoustics; Engineering; Radiology, Nuclear Medicine & Medical Imaging |
| JCDL | Computer Science; Information Science & Library Science |
| KDD | Computer Science |
| MME | Business & Economics; Mathematics; Mathematical Methods In Social Sciences |
| NENE | Environmental Sciences & Ecology; Nuclear Science & Technology |
| ROMAN | Computer Science; Engineering; Robotics |
| SMART | Automation & Control Systems; Computer Science; Transportation |
| WACV | Computer Science; Imaging Science & Photographic Technology; Computer Science |
| WEBCONF | Computer Science |

**Table A3.** Number of conference contributions (conference papers) by conference series and year

|          | 2017 | 2018 | 2019 | 2020 | 2021 | 2022 | 2023 |
|----------|------|------|------|------|------|------|------|
| ACHEMS   |      | 631  | 488  |      | 198  | 334  | 302  |
| ACL      | 323  | 405  | 719  | 511  | 1113 | 1096 | 1133 |
| AEA      | 15   | 117  | 113  | 118  | 121  | 149  | 43   |
| AIED     | 88   | 155  | 131  | 127  | 125  | 211  |      |
| AIES     |      | 76   | 91   | 76   | 114  | 115  | 101  |
| BIOPHYS  | 1007 | 3428 | 2883 | 3052 | 1765 | 2647 |      |
| CHI      | 599  | 1214 | 1300 | 1321 | 873  | 1128 | 879  |
| COMPNET  | 65   | 103  | 123  | 161  |      | 137  | 104  |
| CSEDU    | 155  |      | 148  | 134  | 144  | 164  | 0    |
| DGO      | 89   | 132  | 72   |      |      | 69   | 86   |
| ECGBL    | 125  | 116  | 126  |      | 112  |      |      |
| ECKM     | 156  | 140  | 145  | 119  | 116  |      |      |
| EDUCON   | 291  | 292  | 239  | 303  | 263  | 318  |      |
| EUCAP    | 837  |      | 882  | 934  | 604  | 709  | 876  |
| GLOBECOM | 1017 | 988  | 1086 | 913  | 900  | 1086 |      |
| HEAD     | 161  | 189  | 161  | 163  | 161  | 163  |      |
| ICANN    | 191  | 219  | 323  | 139  | 265  | 259  | 457  |
| ICIP     | 927  | 837  | 943  | 695  | 779  | 844  | 705  |
| ICML     |      | 621  | 774  | 561  | 1185 | 1234 |      |
| IJCNN    | 620  | 755  | 798  | 1122 | 1180 | 1244 | 969  |
| INTER    | 839  | 791  | 951  | 1031 | 987  | 1121 |      |
| IUS      | 578  | 592  | 682  | 542  | 559  | 614  |      |
| JCDL     | 73   | 102  | 115  |      | 72   | 59   |      |
| KDD      | 244  | 306  | 374  | 431  | 478  | 534  | 564  |
| MME      | 150  | 112  | 98   | 103  | 90   | 66   |      |
| NENE     | 117  | 105  | 90   | 112  | 121  |      |      |
| ROMAN    | 231  | 188  | 187  | 209  | 191  | 236  | 355  |
| SMART    | 94   | 122  | 127  | 90   | 84   |      |      |
| WACV     | 152  | 232  | 253  | 402  | 429  | 486  | 707  |
| WEBCONF  | 445  | 572  | 617  | 379  | 487  | 357  | 291  |

**Table A4.** Number of authors of conference contributions by conference series and year

|         | 2017 | 2018  | 2019  | 2020  | 2021 | 2022  | 2023 |
|---------|------|-------|-------|-------|------|-------|------|
| ACHEMS  |      | 3014  | 2221  |       | 1061 | 1907  | 1698 |
| ACL     | 1317 | 1722  | 3295  | 2566  | 5940 | 6189  | 6934 |
| AEA     | 23   | 383   | 345   | 386   | 411  | 465   | 162  |
| AIED    | 402  | 653   | 489   | 565   | 609  | 871   |      |
| AIES    |      | 249   | 292   | 260   | 459  | 347   | 327  |
| BIOPHYS | 4433 | 15716 | 12993 | 14543 | 8404 | 12669 |      |
| CHI     | 2691 | 5423  | 5809  | 6403  | 4027 | 5508  | 4475 |
| COMPNET | 236  | 371   | 435   | 594   |      | 531   | 375  |
| CSEDU   | 561  |       | 567   | 536   | 509  | 602   |      |
| DGO     | 287  | 462   | 249   |       |      | 227   | 310  |
| ECGBL   | 405  | 368   | 355   |       | 356  |       |      |
| ECKM    | 411  | 365   | 383   | 298   | 286  |       |      |
| EDUCON  | 987  | 1084  | 845   | 1110  | 1042 | 1183  |      |
| EUCAP   | 3639 |       | 3866  | 4074  | 2759 | 3189  | 4222 |
| GLOBECOM| 4287 | 4328  | 4889  | 4019  | 4049 | 5169  |      |
| HEAD    | 493  | 568   | 511   | 482   | 494  | 505   |      |
| ICANN   | 740  | 897   | 1436  | 623   | 1217 | 1151  | 2159 |
| ICIP    | 3821 | 3498  | 4083  | 3022  | 3299 | 3758  | 3290 |
| ICML    |      | 2773  | 3495  | 2577  | 5774 | 6483  |      |
| IJCNN   | 2433 | 3143  | 3394  | 4837  | 5276 | 5561  | 4469 |
| INTER   | 3624 | 3541  | 4549  | 5016  | 5164 | 5804  |      |
| IUS     | 3177 | 3356  | 3760  | 2904  | 3017 | 3364  |      |
| JCDL    | 247  | 355   | 379   |       | 297  | 209   |      |
| KDD     | 1118 | 1450  | 1904  | 2229  | 2894 | 3162  | 3500 |
| MME     | 290  | 214   | 180   | 200   | 190  | 128   |      |
| NENE    | 479  | 415   | 301   | 449   | 444  |       |      |
| ROMAN   | 957  | 801   | 845   | 1000  | 855  | 1038  | 1735 |
| SMART   | 398  | 456   | 625   | 397   | 354  |       |      |
| WACV    | 605  | 960   | 1077  | 1814  | 2052 | 2247  | 3444 |
| WEBCONF | 1978 | 2356  | 2666  | 1796  | 2496 | 1854  | 1441 |

**Table A5.** Share of women among the authors of conference contributions by conference series and year

|          | 2017  | 2018  | 2019  | 2020  | 2021  | 2022  | 2023  |
|----------|-------|-------|-------|-------|-------|-------|-------|
| ACHEMS   |       | 43.7% | 41.9% |       | 47.9% | 47.4% | 45.8% |
| ACL      | 26.8% | 26.6% | 31.2% | 28.1% | 31.9% | 33.2% | 33.4% |
| AEA      | 13.0% | 27.4% | 29.9% | 37.0% | 42.1% | 37.1% | 33.8% |
| AIED     | 38.7% | 36.1% | 36.7% | 34.7% | 36.5% | 37.0% |       |
| AIES     |       | 27.9% | 30.4% | 33.3% | 36.9% | 43.1% | 39.4% |
| BIOPHYS  | 32.2% | 32.9% | 34.3% | 34.2% | 35.9% | 35.5% |       |
| CHI      | 35.5% | 36.2% | 37.3% | 39.3% | 40.8% | 41.8% | 40.6% |
| COMPNET  | 24.9% | 23.5% | 22.5% | 25.2% |       | 23.4% | 24.5% |
| CSEDU    | 35.8% |       | 36.7% | 35.2% | 41.4% | 38.9% |       |
| DGO      | 33.2% | 36.6% | 37.2% |       |       | 33.3% | 40.8% |
| ECGBL    | 38.6% | 34.4% | 46.5% |       | 46.8% |       |       |
| ECKM     | 48.8% | 49.7% | 45.6% | 46.6% | 49.4% |       |       |
| EDUCON   | 29.5% | 33.6% | 34.2% | 38.7% | 34.9% | 36.6% |       |
| EUCAP    | 17.2% |       | 18.6% | 18.8% | 16.9% | 18.0% | 16.6% |
| GLOBECOM | 28.6% | 27.7% | 27.9% | 28.9% | 29.4% | 29.5% |       |
| HEAD     | 54.0% | 54.4% | 60.7% | 54.4% | 56.4% | 56.7% |       |
| ICANN    | 21.8% | 23.8% | 22.1% | 25.8% | 31.3% | 32.0% | 37.3% |
| ICIP     | 28.8% | 27.5% | 30.4% | 26.1% | 29.2% | 26.3% | 30.3% |
| ICML     |       | 17.9% | 16.1% | 19.2% | 21.4% | 22.7% |       |
| IJCNN    | 23.5% | 24.1% | 27.4% | 25.3% | 30.8% | 29.2% | 32.7% |
| INTER    | 27.0% | 28.0% | 27.4% | 28.5% | 29.8% | 27.8% |       |
| IUS      | 23.6% | 25.1% | 24.4% | 26.1% | 31.4% | 26.8% |       |
| JCDL     | 29.8% | 27.2% | 30.9% |       | 26.8% | 27.6% |       |
| KDD      | 26.3% | 30.3% | 30.7% | 30.4% | 32.4% | 30.7% | 32.5% |
| MME      | 37.1% | 32.5% | 40.4% | 36.5% | 42.1% | 40.8% |       |
| NENE     | 13.5% | 14.6% | 12.6% | 14.4% | 17.5% |       |       |
| ROMAN    | 27.1% | 31.0% | 28.4% | 30.5% | 33.7% | 37.9% | 34.5% |
| SMART    | 27.0% | 26.9% | 24.5% | 26.7% | 29.7% |       |       |
| WACV     | 25.7% | 22.0% | 18.6% | 21.0% | 21.5% | 22.0% | 21.7% |
| WEBCONF  | 24.6% | 26.7% | 29.6% | 32.3% | 30.6% | 32.9% | 30.5% |

**Table A6.** Share of women among the authors of conference contributions in the sample by country of affiliation (only the top 60 countries with the highest number of authors are presented)

| Country | Share of women |
| --- | --- |
| Arab Emirates | 26.3% |
| Argentina | 46.4% |
| Australia | 32.1% |
| Austria | 29.3% |
| Bangladesh | 29.8% |
| Belgium | 24.6% |
| Brazil | 23.0% |
| Canada | 29.4% |
| Chile | 22.0% |
| China | 40.5% |
| Colombia | 38.7% |
| Croatia | 27.4% |
| Cyprus | 23.0% |
| Czech Republic | 26.9% |
| Denmark | 27.6% |
| Ecuador | 43.3% |
| Egypt | 26.2% |
| Estonia | 37.8% |
| Finland | 21.7% |
| France | 26.0% |
| Germany | 22.5% |
| Greece | 16.5% |
| Hungary | 25.8% |
| India | 24.7% |
| Iran | 25.5% |
| Ireland | 28.5% |
| Israel | 23.6% |
| Italy | 28.4% |
| Japan | 15.2% |
| Lithuania | 40.9% |
| Luxembourg | 28.0% |
| Malaysia | 41.7% |
| Mexico | 35.9% |
| Morocco | 31.0% |
| Netherlands | 30.5% |
| New Zealand | 30.4% |
| Norway | 27.2% |
| Pakistan | 19.2% |
| Peru | 19.5% |
| Poland | 25.5% |
| Portugal | 33.4% |
| Qatar | 15.0% |
| Romania | 32.1% |
| Russia | 29.4% |
| Saudi Arabia | 18.2% |
| Singapore | 31.8% |
| Slovakia | 30.3% |
| Slovenia | 12.1% |
| South Africa | 35.7% |
| South Korea | 22.0% |
| Spain | 30.0% |
| Sweden | 31.2% |
| Switzerland | 22.4% |
| Taiwan | 47.4% |
| Thailand | 35.5% |
| Tunisia | 37.5% |
| Turkey | 33.2% |
| United Kingdom | 28.8% |
| USA | 32.9% |
| Vietnam | 60.9% |